\documentstyle[aps,prd]{revtex}
\begin{document}
\draft
\title{Birefringent Electroweak Textures}
\author{Marcus J. Thatcher
\thanks{Email address: Marcus.Thatcher@sci.monash.edu.au; Fax: +61 3 9905 3637.}
and Michael J. Morgan
\thanks{Email address: Michael.Morgan@sci.monash.edu.au; Fax: +61 3 9905 3637.}
}
\address{Department of Physics, Monash University, Clayton 3168, Victoria, Australia}
\maketitle
\begin{abstract}
The behaviour of electromagnetic waves propagating through an electroweak homilia
string network is examined. This string network is topologically stable as a cosmic
texture, and is characterized by the spatial variation of the isospin
rotation of the Higgs field.
As a consequence the photon field couples to the intermediate
vector bosons, producing a finite range electromagnetic field. It is
found that the propagation speed of the photon depends on its polarization
vector, whence an homilia string network acts as a birefringent medium.  We
estimate the birefringent scale for this texture and show that it
depends on the frequency of the electromagnetic wave and the length scale
of the homilia string network.
\end{abstract}
\pacs{PACS numbers: 98.80.Cq,11.27.+d,11.15.-q}

\section{Introduction}
In the gauge theory of electroweak interactions
the massless photon and the massive intermediate vector bosons,
$Z^{0}$ and $W^{\pm}$, are unified within the group $SU(2)\times U(1)$
\cite{greiner93}. 
Recently we have shown that the electroweak model allows for
the formation of a new kind of defect, called a homilia string \cite{thatcher99}.
Although locally string-like, this defect is homotopicaly classified as a texture.
Since an homilia network behaves like a string locally,
it is governed by the dynamics of a string network, thereby avoiding
collapse associated with
spherical texture defects (see for example \cite{vilenkin94}). Moreover,
the local string-like geometry of homilia strings leads to a non-zero
energy density that distinguishes homilia strings from the vacuum
(i.e., $(D_{\mu}\Phi)^{\dagger} D^{\mu}\Phi$ and $Tr(F_{\mu\nu}F^{\mu\nu})$
are non-zero). As with textures and non-symmetric strings
\cite{axenides97}, homilia strings reconcile an undefined phase by forcing the Higgs field to
undergo a rotation in isospin space, rather than forming a region of
false vacuum.

The behaviour of photons in regions where the Higgs field varies spatially
has been discussed previously \cite{nambu77,vachaspati91,tornkvist98}.
Nambu \cite{nambu77}
and Vachaspati \cite{vachaspati91} proposed definitions of the
electromagnetic field tensor which
lead to a finite range electromagnetic field within some topological defects (e.g., sphalerons \cite{klinkhamer84}).
However, a finite range photon field
is considered to be unphysical in this context
and Tornkvist \cite{tornkvist98} has proposed an ansatz intended to preserve
the power-law behaviour of electromagnetic fields.

In this paper we examine the behaviour of electromagnetic waves propagating
through an electroweak homilia string network. This network is characterized by
the spatial variation of the orientation of the Higgs field
in isospin space. The results reported here differ from previous work, since
we analyse the behaviour of propagating electromagnetic waves when the
definition of the vector bosons is endowed with an explicit spatial dependence.
It is found that
photons couple to massive intermediate vector bosons when the
Higgs field rotates in isospin space. In this situation
it is not possible to solve for propagating electromagnetic
waves (photons) independently of the massive vector bosons, and all propagating solutions
consist of a mixture of photons, $Z^{0}$ and $W^{\pm}$ particles. Since
photons propagate in the company of massive vector bosons, the phase velocity
of electromagnetic waves
is reduced. As a consequence cosmic textures, such as homilia string networks,
have an effective refractive index which depends
on the polarization state of the photon. These textures act like
a birefringent medium and might be observable via measurements of cosmological
anisotropy.

This paper is organised as follows. Section \ref{sec-hgauge} describes
the choice of gauge which is utilised to simplify the analysis of electroweak textures.
However, the results are not dependent on the choice of gauge, and 
where relevant we demonstrate gauge invariance of the model predictions.
In Sec.\ \ref{sec-eigenvectors}
the vector boson eigenvectors are calculated from a spatially dependent
mass matrix. The coupling of the photon field to massive vector bosons
is examined in Sec.\ \ref{sec-coupling}. In Sec.\ \ref{sec-isospinanalytic}
we solve for the propagating solutions
of bosons fields using the coupled equations of motion. Finally, in Sec.\ \ref{sec-measure}
we estimate the strength of the isospin coupling
and the birefringence length scale using a
homilia string network model. In appendix \ref{sec-gauge} we calculate the
isospin coupling for an arbitrary choice of gauge and
demonstrate gauge invariance of the model.

\section{Homilia strings and gauge transformations}
\label{sec-hgauge}
Homilia strings were introduced in a earlier paper \cite{thatcher99}, where
the orientation of the Higgs field in isospin space was chosen so that homilia strings
separate into distinct components of the Higgs isodoublet. This enables
us to define an $\alpha$-string and a $\beta$-string within the context of 
the so called homilia gauge \cite{thatcher99}.
The homilia $\alpha$-string is described in the homilia gauge as
\begin{equation}
\label{eq-homilia}
\Phi^{\alpha}_{H} = \left(\begin{array}{c} |\phi^{\alpha}(r)| e^{i\theta} \\ |\phi^{\beta}(r)| \end{array} \right),
\end{equation}
where $|\phi^{\alpha}(r)|$ and $|\phi^{\beta}(r)|$ are the real magnitudes
of the scalar fields and the polar co-ordinate, $\theta$, describes the non-trivial phase winding.  The solutions
and boundary conditions for $|\phi^{\alpha}(r)|$ and $|\phi^{\beta}(r)|$ are
discussed in detail in reference \cite{thatcher99}.

To simplify the following discussion we perform a local gauge
transformation
\begin{equation}
\label{eq-gaugetrans}
\Phi^{\alpha}_{H} \rightarrow \Phi^{\alpha} = e^{-i\tau^{\alpha}\theta} \Phi^{\alpha}_{H} = \left(\begin{array}{c} |\phi^{\alpha}(r)| \\ |\phi^{\beta}(r)| \end{array}\right),
\end{equation}
where $\tau^{\alpha}$ is the $\alpha$-string generator
\begin{equation}
\tau^{\alpha}=\left[\begin{array}{cc} 1 & 0 \\ 0 & 0 \end{array}\right].
\end{equation}
The gauge transformation (\ref{eq-gaugetrans}) results in a real
Higgs field. It is important to emphasize that the line of
undefined phase cannot be gauged away, since the gauge transformation is
not defined at $r=0$ (i.e., Eq.\ (\ref{eq-gaugetrans})
requires $|\phi^{\alpha}(r=0)|=0$, which preserves
the undefined phase). Similarly a
local gauge transformation can be performed for
$\beta$-strings, using the $\beta$-string generator $\tau^{\beta}$, where
\begin{equation}
\tau^{\beta}=\left[\begin{array}{cc} 0 & 0 \\ 0 & 1 \end{array}\right].
\end{equation}
We stress that the gauge transformation in Eq.\ (\ref{eq-gaugetrans}) is
not essential to the following discussion, but has been introduced for simplicity
since it results in a real Higgs field. In appendix \ref{sec-gauge} we
demonstrate that the results are valid for an arbitrary choice of gauge.

\section{Eigenvectors of the Mass Matrix}
\label{sec-eigenvectors}
The Lagrangian density that describes the $SU(2)\times U(1)$ electroweak
model is written as
\begin{equation}
\label{eq-ewlag}
{\cal L} = (D^{\mu}\Phi)^{\dagger}(D_{\mu}\Phi)-\frac{1}{4}Tr(F_{\mu\nu}F^{\mu\nu})-\frac{1}{4}\lambda(\Phi^{\dagger}\Phi-\eta^{2})^{2},
\end{equation}
where $\Phi$ is a complex isodoublet,
$D_{\mu}=\partial_{\mu}-iqA_{\mu}$, $F_{\mu\nu}=\partial_{\mu}A_{\nu}-\partial_{\nu}A_{\mu}+iq[A_{\mu},A_{\nu}]$,
$A_{\mu}=A^{a}_{\mu}\sigma_{a}$ and $\sigma_{a}$ ($a=1,2,3$) denotes the Pauli
spin matrices. To include the $U(1)$ symmetry we have introduced the field $A^{0}_{\mu}$,
which couples to the $U(1)$ generator via $\sigma_{0}\equiv I \tan\theta_{W}$.
The Lagrangian density (\ref{eq-ewlag}) can be expanded
to obtain
\begin{eqnarray}
{\cal L} & = & (\partial^{\mu}\Phi)^{\dagger}(\partial_{\mu}\Phi)+iqA^{\mu}_{a}\left[(\sigma^{a}\Phi)^{\dagger}\partial_{\mu}\Phi-(\partial_{\mu}\Phi)^{\dagger}\sigma^{a}\Phi\right] 
+ q^{2} A_{\mu}^{a}A^{\mu}_{b} \Phi^{\dagger} [\sigma_{a},\sigma^{b}]_{+}\Phi \nonumber \\
& & - \frac{1}{2}\left(\partial_{\mu}A_{\nu}^{a}-\partial_{\nu}A_{\mu}^{a}+2q\epsilon^{a}_{bc}A^{b}_{\mu}A^{c}_{\nu}\right)
\left(\partial^{\mu}A^{\nu}_{a}-\partial^{\nu}A^{\mu}_{a}+2q\epsilon_{a}^{bc}A^{\mu}_{b}A^{\nu}_{c}\right) \nonumber \\
\label{eq-ewlagexp}
& & - \frac{1}{4}\lambda\left(\Phi^{\dagger}\Phi-\eta^{2}\right)^{2},
\end{eqnarray}
where $[\sigma_{a},\sigma_{b}]_{+}$ represents the anitcommutation relation,
$\sigma_{a}\sigma_{b}+\sigma_{b}\sigma_{a}$.
Equation (\ref{eq-ewlagexp}) describes the dynamics of the Higgs
field $\Phi$, the vector gauge fields $A_{\mu}^{a}$ and the interaction
between the Higgs field and gauge fields.
The equations of motion for the vector gauge fields are
\begin{equation}
\label{eq-eweom}
\mbox{Tr}\left(\sigma^{a\,2}\right)\left\{\partial^{\mu}F_{\mu\nu}^{a}-\frac{q}{2} f^{a}_{bc}A^{b\,\mu}F^{c}_{\mu\nu}\right\}-i\frac{q}{2}\left\{\left(D_{\nu}\Phi\right)^{\dagger}\sigma^{a}\Phi-\left(\sigma^{a}\Phi\right)^{\dagger}D_{\nu}\Phi\right\} = 0.
\end{equation}
Equation (\ref{eq-eweom}) can be expanded to obtain
\begin{eqnarray}
\Box A_{\nu}^{a} - \partial_{\nu} \partial^{\mu} A_{\mu}^{a} &= & -\frac{q^{2}}{4} (\Phi^{\dagger} [\sigma_{b},\sigma^{a}]_{+} \Phi ) A_{\nu}^{b} - \frac{q}{2}f^{a}_{bc}A^{b\,\mu}\partial_{\nu}A^{c}_{\mu}- \frac{1}{2}f^{a}_{bc}A^{c}_{\nu}\partial^{\mu}A_{\mu}^{b} \nonumber \\
\label{eq-gaugeeombreakdown}
& & + \frac{q^{2}}{4} (f^{a}_{bc} f^{c}_{de} A^{b\,\mu}  A^{d}_{\mu}A^{e}_{\nu}) + i \frac{q}{2} [(\partial_{\nu}\Phi)^{\dagger} \sigma^{a} \Phi - (\sigma^{a} \Phi)^{\dagger}\partial_{\nu}\Phi].
\end{eqnarray}

The last term in Eq.\ (\ref{eq-gaugeeombreakdown}) is a current source term.
Topological defects, such as homilia strings \cite{thatcher99},
textures and sphalerons \cite{klinkhamer84},
describe variations in the Higgs field which result in a non-zero
current term in Eq.\ (\ref{eq-gaugeeombreakdown}).
For homilia strings (HS) the current
term leads to static (non-trivial) vector boson fields described by
\begin{equation}
\label{eq-staticgf}
A_{\theta}^{HS} = \frac{n}{e r}\left(\begin{array}{cc} b(r) & c(r)e^{-in\theta} \\ c(r)e^{in\theta} & d(r) \end{array}\right).
\end{equation}
Applying the gauge transformation (\ref{eq-gaugetrans}) to Eq.\ (\ref{eq-staticgf})
results in
\begin{equation}
A_{\theta}^{HS} = \frac{n}{e r}\left(\begin{array}{cc} b(r)-1 & c(r) \\ c(r) & d(r) \end{array}\right).
\end{equation}
The functions $b(r)$, $c(r)$ and $d(r)$ have been calculated numerically in
reference \cite{thatcher99}. Here we are interested in
the implications arising from the spatial dependence
of the vector boson eigenvectors, rather than the effects due to the current
term. Consequently, in what follows we neglect the contribution from the
current term.

Consider the mass matrix constructed from the $SU(2)\times U(1)$ Lagrangian
density. From Eq.\ (\ref{eq-ewlagexp}), the mass matrix term is
\begin{equation}
q^{2} A_{\mu}^{a}A^{\mu}_{b} \Phi^{\dagger} [\sigma_{a},\sigma^{b}]_{+}\Phi.
\end{equation}
We define
\begin{equation}
{\cal M}_{ab} = q^{2} \Phi^{\dagger} [\sigma_{a},\sigma_{b}]_{+} \Phi,
\end{equation}
which is written explicitly in matrix form as
\begin{equation}
\label{eq-massmatrix}
{\cal M} =\left(\begin{array}{cccc} {\cal M}_{00} & {\cal M}_{01} & {\cal M}_{02} & {\cal M}_{03} \\ {\cal M}_{10} & {\cal M}_{11} &  & \vdots \\ {\cal M}_{20} & & \ddots & \\ {\cal M}_{30} & \cdots & & {\cal M}_{33} \end{array}\right)
= q^{2}\left(\begin{array}{cccc} |\Phi|^{2} \tan^{2}\theta_{W} & \chi^{1}(x) & \chi^{2}(x) & \chi^{3}(x) \\ \chi^{1}(x) & |\Phi|^{2} & 0 & 0 \\ \chi^{2}(x) & 0 & |\Phi|^{2} & 0 \\ \chi^{3}(x) & 0 & 0 & |\Phi|^{2} \end{array}\right),
\end{equation}
where $\theta_{W}$ is the Weinberg angle and $|\Phi|^{2}=|\phi^{\alpha}|^{2}+|\phi^{\beta}|^{2}$.
The $\chi^{a}(x)$-terms describe the isospin rotation of the Higgs field
and are defined by
\begin{equation}
\chi^{a}(x)=\tan\theta_{W}\Phi^{\dagger}(x)\sigma^{a}\Phi(x)\,\,\,\,\,\,\,\,(a=1,2,3),
\end{equation}
for which the following identity is valid
\begin{equation}
\label{eq-isoidenity}
|\Phi|^{4}\tan^{2}\theta_{W} \equiv (\chi^{1}(x))^{2}+ (\chi^{2}(x))^{2} + (\chi^{3}(x))^{2}.
\end{equation}

Consider the situation
where $|\Phi|$ is constant, but $\phi^{\alpha}$ and $\phi^{\beta}$ vary spatially.
The Higgs field is written as
\begin{equation}
\Phi({\bf x}) = \left(\begin{array}{c} \phi^{\alpha}({\bf x}) \\ \phi^{\beta}({\bf x}) \end{array}\right).
\end{equation}
To simplify the analysis we perform a
gauge transformation (\ref{eq-gaugetrans}). This results in a
real Higgs field, with $\chi^{2}=0$. The Higgs field now becomes
\begin{equation}
\Phi({\bf x}) = \left(\begin{array}{c} |\phi^{\alpha}({\bf x})| \\ |\phi^{\beta}({\bf x})| \end{array}\right).
\end{equation}
The independent fields are determined from the mass matrix ${\cal M}$.
The vector boson eigenvectors of the mass matrix (\ref{eq-massmatrix}) are (with $\chi^{2}=0$)
\begin{mathletters}
\label{eq-gaugeeigenvectors}
\begin{eqnarray}
\label{eq-gaugeeigenvectorsbeg}
{\bf A}^{\gamma}(x) & = &\frac{\sin\theta_{W}}{|\Phi|^{2}\tan\theta_{W}} \left(\begin{array}{c} -|\Phi|^{2} \\ \chi^{1}(x) \\ 0 \\ \chi^{3}(x) \end{array}\right) \\
{\bf A}^{Z}(x) & = &\frac{\cos\theta_{W}}{|\Phi|^{2}\tan\theta_{W}} \left(\begin{array}{c} |\Phi|^{2}\tan^{2}\theta_{W} \\ \chi^{1}(x) \\ 0 \\ \chi^{3}(x) \end{array}\right) \\
{\bf A}^{W}(x) & = &\frac{1}{|\Phi|^{2}\tan\theta_{W}}\left(\begin{array}{c} 0\\ \chi^{3}(x) \\ 0 \\ -\chi^{1}(x) \end{array}\right) \\
{\bf A}^{\overline{W}} & = & \left(\begin{array}{c} 0\\ 0 \\ i \\ 0 \end{array}\right).
\end{eqnarray}
\end{mathletters}
Here $A^{\gamma}$ and $A^{Z}$ represent the photon and $Z^{0}$ eigenvectors, respectively.
The vector bosons $W^{+}$ and $W^{-}$ are described by the
linear combinations
\begin{mathletters}
\begin{eqnarray}
A^{W} & = & \frac{1}{2}\left(A^{W^{+}} + A^{W^{-}}\right) \\ 
A^{\overline{W}} & = & \frac{1}{2}\left(A^{W^{+}} - A^{W^{-}}\right).
\end{eqnarray}
\end{mathletters}
$W^{+}$ and $W^{-}$ are defined to
be consistent with the standard decomposition of the gauge particles
in the limit of the electroweak gauge \cite{greiner93,leader96,weinberg96}.
Note that the eigenvectors (\ref{eq-gaugeeigenvectors}) now depend on the space
co-ordinate, which follows from the spatial dependence of the
orientation of the Higgs field.

If we do not perform a local gauge transformation (\ref{eq-gaugetrans}), the
photon eigenvector becomes
\begin{equation}
\label{eq-photoncomp}
{\bf A}^{\gamma}(x) = \frac{\sin\theta_{W}}{|\Phi|^{2}\tan\theta_{W}} \left(\begin{array}{c} -|\Phi|^{2} \\ \chi^{1}(x) \\ \chi^{2}(x) \\ \chi^{3}(x) \end{array}\right).
\end{equation}
The photon field defined in Eq.\ (\ref{eq-photoncomp}) is equivalent to
the definition in reference \cite{vachaspati91}. However, unlike the definitions
utilised in references \cite{nambu77,vachaspati91,tornkvist98}
we do not explicitly define the electromagnetic field tensor. Rather, we examine
the behaviour of photons (\ref{eq-photoncomp}) in
the context of the $SU(2)\times U(1)$ electroweak unification model.
In Sec.\ \ref{sec-isospinanalytic}, we show that
propagating solutions involve a mixture of vector bosons, and hence
electromagnetic waves do not propagate independently of the massive
vector bosons. The gauge invariance of the combined (propagating) state
follows as a consequence of the gauge invariance of $Tr(F_{\mu\nu}F^{\mu\nu})$
(see reference \cite{greiner93}).

\section{Isospin Rotation and Vector Bosons}
\label{sec-coupling}
The spatial dependence of the photon eigenvector, ${\bf A}^{\gamma}(x)$,
in Eq.\ (\ref{eq-gaugeeigenvectorsbeg}) is a consequence
of $\Phi$ undergoing isospin rotation. This results in the photon field
coupling to the intermediate vector bosons.
To understand the consequences of this spatial dependence,
we write the gauge fields in terms of the
eigenvectors (\ref{eq-gaugeeigenvectors}). Thus, for $\chi^{2}=0$
\begin{mathletters}
\label{eq-gaugedef}
\begin{eqnarray}
A^{0}_{\mu} & = & -\cos\theta_{W} \alpha^{\gamma}_{\mu} + \sin\theta_{W} \alpha^{Z}_{\mu}\\
A^{1}_{\mu} & = & \frac{1}{|\Phi|^{2}\tan\theta_{W}}\left(\sin\theta_{W}\chi^{1} \alpha^{\gamma}_{\mu} + \cos\theta_{W}\chi^{1} \alpha^{Z}_{\mu} + \chi^{3} \alpha^{W}_{\mu}\right) \\
A^{2}_{\mu} & = & i\alpha^{\overline{W}}_{\mu} \\
A^{3}_{\mu} & = & \frac{1}{|\Phi|^{2}\tan\theta_{W}}\left(\sin\theta_{W}\chi^{3} \alpha^{\gamma}_{\mu} + \cos\theta_{W}\chi^{3} \alpha^{Z}_{\mu} - \chi^{1} \alpha^{W}_{\mu}\right).
\end{eqnarray}
\end{mathletters}
The fields $\alpha^{\gamma}_{\mu}$, $\alpha^{Z}_{\mu}$, $\alpha^{W}_{\mu}$
and $\alpha^{\overline{W}}_{\mu}$ define the
boson fields for an arbitrary isospin orientation of the Higgs field.
Substituting Eqs.\ (\ref{eq-gaugedef})
into the Yang-Mills field tensor results in
\begin{eqnarray}
{\cal L}_{YM} & = & Tr(F_{\mu\nu}F^{\mu\nu}) \nonumber \\
& = & (\partial_{\mu}\alpha^{\gamma}_{\nu} - \partial_{\nu}\alpha^{\gamma}_{\mu})(\partial^{\mu}\alpha^{\gamma\,\nu} -\partial^{\nu}\alpha^{\gamma\,\mu})
+(\partial_{\mu}\alpha^{Z}_{\nu} - \partial_{\nu}\alpha^{Z}_{\mu})(\partial^{\mu}\alpha^{Z\,\nu} -\partial^{\nu}\alpha^{Z\,\mu}) \nonumber \\
& &+(\partial_{\mu}\alpha^{W}_{\nu} - \partial_{\nu}\alpha^{W}_{\mu})(\partial^{\mu}\alpha^{W\,\nu} -\partial^{\nu}\alpha^{W\,\mu})
+(\partial_{\mu}\alpha^{\overline{W}}_{\nu} - \partial_{\nu}\alpha^{\overline{W}}_{\mu})(\partial^{\mu}\alpha^{\overline{W}\,\nu} -\partial^{\nu}\alpha^{\overline{W}\,\mu}) \nonumber \\
& &+2\rho^{\mu}\alpha^{\gamma\,\nu}\sin^{2}\theta_{W}\left(\rho_{\mu}\alpha^{\gamma}_{\nu} - \rho_{\nu}\alpha^{\gamma}_{\mu}\right)
+2\rho^{\mu}\alpha^{Z\,\nu}\cos^{2}\theta_{W}\left(\rho_{\mu}\alpha^{Z}_{\nu} - \rho_{\nu}\alpha^{Z}_{\mu}\right) \nonumber \\
& &+2\rho^{\mu}\alpha^{W\,\nu}\left(\rho_{\mu}\alpha^{W}_{\nu} - \rho_{\nu}\alpha^{W}_{\mu}\right)
+2\rho^{\mu}\alpha^{\gamma\,\nu}\sin\theta_{W}\cos\theta_{W}\left(\rho_{\mu}\alpha^{Z}_{\nu} - \rho_{\nu}\alpha^{Z}_{\mu}\right) \nonumber \\
& &+2\rho^{\mu}\alpha^{\gamma\,\nu}\sin\theta_{W}\left(\partial_{\mu}\alpha^{W}_{\nu} - \partial_{\nu}\alpha^{W}_{\mu}\right)
+2\rho^{\mu}\alpha^{Z\,\nu}\cos\theta_{W}\left(\partial_{\mu}\alpha^{W}_{\nu} - \partial_{\nu}\alpha^{W}_{\mu}\right) \nonumber \\
& & -2\rho^{\mu}\alpha^{W\,\nu}\sin\theta_{W}\left(\partial_{\mu}\alpha^{\gamma}_{\nu} - \partial_{\nu}\alpha^{\gamma}_{\mu}\right)
-2\rho^{\mu}\alpha^{W\,\nu}\cos\theta_{W}\left(\partial_{\mu}\alpha^{Z}_{\nu} - \partial_{\nu}\alpha^{Z}_{\mu}\right) \nonumber \\
\label{eq-ewlagfullterms}
& &+{\cal L}_{int}(q,q^2)+{\cal L}_{int+}(q \rho_{\mu}),
\end{eqnarray}
where ${\cal L}_{int}(q,q^2)$ represents terms involving the
structure constants (see reference \cite{greiner93}),
and ${\cal L}_{int+}(q \rho_{\mu})$ includes terms 
involving a mixture
of $q$ and the strength of the isospin rotation, $\rho_{\mu}$, i.e.,
\begin{eqnarray}
{\cal L}_{int+}(q \rho_{\mu})& =& i 2 q \alpha^{\gamma\,\nu}\rho^{\mu}\sin^{2}\theta_{W}(\alpha^{\overline{W}}_{\mu}\alpha^{\gamma}_{\nu}-\alpha^{\gamma}_{\mu}\alpha^{\overline{W}}_{\nu}) \nonumber \\
&&+i 2 q \alpha^{\gamma\,\nu}\rho^{\mu} \sin\theta_{W}\cos\theta_{W}(\alpha^{\overline{W}}_{\mu}\alpha^{Z}_{\nu}-\alpha^{Z}_{\mu}\alpha^{\overline{W}}_{\nu}) \nonumber \\
&&+ i 2 q \alpha^{Z\,\nu}\rho^{\mu}\cos^{2}\theta_{W}(\alpha^{\overline{W}}_{\mu}\alpha^{Z}_{\nu}-\alpha^{Z}_{\mu}\alpha^{\overline{W}}_{\nu}) \nonumber \\
&&+i 2 q \alpha^{Z\,\nu}\rho^{\mu}\sin\theta_{W}\cos\theta_{W} (\alpha^{\overline{W}}_{\mu}\alpha^{\gamma}_{\nu}-\alpha^{\gamma}_{\mu}\alpha^{\overline{W}}_{\nu}) \nonumber \\
&&+ i 2 q \alpha^{W\,\nu}\rho^{\mu}(\alpha^{\overline{W}}_{\mu}\alpha^{W}_{\nu}-\alpha^{W}_{\mu}\alpha^{\overline{W}}_{\nu}).
\end{eqnarray}

The first four terms in Eq.\ (\ref{eq-ewlagfullterms})
are the usual field strength terms. However, the remaining terms are a
consequence of the spatial dependence of $\chi^{a}(x)$.
The strength of the isospin rotation, $\rho_{\mu}$, is defined as
\begin{equation}
\label{eq-rhodefa}
\rho_{\nu}= \frac{(\partial_{\nu}\chi^{1})\chi^{3}-(\partial_{\nu}\chi^{3})\chi^{1}}{|\Phi|^{4}\tan^{2}\theta_{W}} \,\,\,\,\,\,\,\, (\chi^{2}=0).
\end{equation}
From Eq.\ (\ref{eq-rhodefa}) we have the relation
\begin{equation}
\rho_{\mu}\rho^{\mu} \equiv \frac{(\partial_{\mu} \chi^{1})(\partial^{\mu}\chi^{1})+(\partial_{\mu}\chi^{3})(\partial^{\mu}\chi^{3})}{|\Phi|^{4}\tan^{2}\theta_{W}}.
\end{equation}
The quantity $\rho_{\mu}$
represents the fundamental coupling strength, since we can write
\begin{mathletters}
\label{eq-fcoup}
\begin{eqnarray}
(\sin\theta_{W})^{-1}\partial_{\mu} {\bf A}^{\gamma} & = & (\cos\theta_{W})^{-1}\partial_{\mu} {\bf A}^{Z} = \rho_{\mu} {\bf A}^{W} \\
\partial_{\mu} {\bf A}^{W} & = & -\rho_{\mu}\left({\bf A}^{\gamma}\sin\theta_{W} + {\bf A}^{Z}\cos\theta_{W}\right).
\end{eqnarray}
\end{mathletters}
Therefore $\rho_{\mu}$ defines the relationship between the isospin rotation
and the derivatives of the eigenvectors of the vector bosons.
From Eqs.\ (\ref{eq-fcoup}) we see that the field strength term,
$Tr(F_{\mu\nu}F^{\mu\nu})$, leads immediately to isospin coupling, as a
direct consequence of the spatial dependence of the eigenvectors.

From Eq.\ (\ref{eq-ewlagfullterms}) we note that the rotation of the Higgs field
produces a term of the form
\begin{equation}
\label{eq-ewphotonpm}
2\rho^{\mu}\alpha^{\gamma\,\nu} \sin^{2}\theta_{W}\left(\rho_{\mu}\alpha^{\gamma}_{\nu} - \rho_{\nu}\alpha^{\gamma}_{\mu}\right).
\end{equation}
If we define the isospin rotation direction to be in the $z$-direction, such
that $\rho_{z}$ is the only non-zero component of $\rho_{\mu}$, then
Eq.\ (\ref{eq-ewphotonpm}) becomes
\begin{equation}
\label{eq-ewphotonpmb}
2\rho_{z}\rho^{z}\sin^{2}\theta_{W}\left( \alpha_{\nu}^{\gamma}\alpha^{\gamma\,\nu}-\alpha_{z}^{\gamma}\alpha^{\gamma\,z}\right).
\end{equation}
When $\nu\neq z$,
Eq.\ (\ref{eq-ewphotonpmb}) represents a pseudo-mass term for the photon
(i.e., when the polarization state is perpendicular to
the $z$-direction). We define the photon pseudo-mass, $P_{\gamma}$, by
\begin{equation}
\label{eq-pseudom}
P_{\gamma}^{2} = \rho_{\mu}\rho^{\mu} \sin^{2}\theta_{W}.
\end{equation}
It is apparent that the rotation of the Higgs field (in isospin space) results
in the photon coupling to the Higgs field.
Coupling of the photon field to the
massive intermediate vector bosons prevents the photon propagating as a
massless particle.
However, the photon rest mass is determined by the eigenvalue of the mass
matrix and is always zero.
If the photon polarization state is parallel to the isospin rotation
(i.e., the $z$-direction), the photon decouples from the massive intermediate
vector bosons and does indeed propagate as a massless particle.
As a consequence of this polarisation dependence
an homilia string network (texture) 
acts as a birefringent medium. In Sec.\ \ref{sec-isospinanalytic} we
examine the consequences of this observation in more detail.

\section{Consequences of Isospin Rotation}
\label{sec-isospinanalytic}
To simplify the analysis of the photon's behaviour in a region of
isospin rotation, we approximate the isospin rotation by an average value
\begin{equation}
\label{eq-aveapprox}
<\rho_{\mu}\rho^{\mu}>=\rho^{2},
\end{equation}
where $\rho$ is a constant. A co-ordinate system is chosen so that the
isospin rotation is in the $z$-direction
\begin{equation}
\rho_{\nu} = \left\{ \begin{array}{cl} \rho & \nu=z \\ 0 & \nu\neq z. \end{array}\right. 
\end{equation}

Adopting the simplification (\ref{eq-aveapprox}), the equations of motion become
\begin{mathletters}
\label{eq-vectoreom}
\begin{eqnarray}
\partial^{\mu}\partial_{\mu}\alpha^{\gamma}_{\nu}-\partial_{\nu}\partial^{\mu}\alpha_{\mu}^{\gamma}
- \sin^{2}\theta_{W}\left[(1-\delta_{\nu z})\rho^{2} (\alpha^{\gamma}_{\nu}+\alpha^{Z}_{\nu})+(2-\delta_{\nu z})\rho \partial_{z}\alpha^{W}_{\nu}\right] &&\nonumber \\
+ \sin^{2}\theta_{W}\left(\rho \delta_{\nu z} \partial^{\mu}\alpha_{\mu}^{W} \right) + {\cal O}(q) & = & 0 \\
\partial^{\mu}\partial_{\mu}\alpha^{Z}_{\nu}-\partial_{\nu}\partial^{\mu}\alpha_{\mu}^{Z}-M_{Z}^{2}\alpha^{Z}_{\nu}
- \cos^{2}\theta_{W}\left[(1-\delta_{\nu z})\rho^{2} (\alpha^{\gamma}_{\nu}+\alpha^{Z}_{\nu})\right] & & \nonumber \\
- \cos^{2}\theta_{W}\left[(2-\delta_{\nu z})\rho \partial_{z}\alpha^{W}_{\nu}-\rho \delta_{\nu z} \partial^{\mu}\alpha^{W}_{\mu}\right]+ {\cal O}(q) & = & 0 \\
\partial^{\mu}\partial_{\mu}\alpha^{W}_{\nu}-\partial_{\nu}\partial^{\mu}\alpha_{\mu}^{W}-M_{W}^{2}\alpha^{W}_{\nu}
-\left[(1-\delta_{\nu z})\rho^{2} \alpha^{W}_{\nu}+(2-\delta_{\nu z})\rho \partial_{z}(\alpha^{\gamma}_{\nu}+\alpha^{Z}_{\nu})\right] && \nonumber \\
+\left[\rho \delta_{\nu z} \partial^{\mu}(\alpha^{\gamma}_{\mu}+\alpha^{Z}_{\mu}) \right] + {\cal O}(q) & = & 0 \\
\label{eq-overweom}
\partial^{\mu}\partial_{\mu}\alpha^{\overline{W}}_{\nu}-\partial_{\nu}\partial^{\mu}\alpha_{\mu}^{\overline{W}}-M_{W}^{2}\alpha^{\overline{W}}_{\nu} + {\cal O}(q) & = & 0,
\end{eqnarray}
\end{mathletters}
where ${\cal O}(q)$ denotes higher order interaction terms,
and $M_{Z}$ and $M_{W}$ are the masses of the $Z^{0}$ and $W^{\pm}$ particles,
respectively, i.e.,
\begin{mathletters}
\begin{eqnarray}
M_{Z}^{2} & = & q^{2}[1+\tan^{2}\theta_{W}]|\Phi|^{2} \\
M_{W}^{2} & = & q^{2}|\Phi|^{2}.
\end{eqnarray}
\end{mathletters}
When analysing the behaviour of the photon, $\alpha^{\overline{W}}(x)$ is set
to zero. We adopt the following ansatzen
for $\alpha^{\gamma}_{\nu}(t,{\bf x})$, $\alpha^{Z}_{\nu}(t,{\bf x})$ and $\alpha^{W}_{\nu}(t,{\bf x})$:
\begin{mathletters}
\label{eq-photonresult}
\begin{eqnarray}
\alpha^{\gamma}_{\nu}(t,{\bf x}) & = & a_{\nu} e^{i({\bf k}\cdot{\bf x}\pm \omega t)} \\
\alpha^{Z}_{\nu}(t,{\bf x}) & = & b_{\nu} e^{i({\bf k}\cdot{\bf x}\pm \omega t)} \\
\alpha^{W}_{\nu}(t,{\bf x}) & = & c_{\nu} e^{i({\bf k}\cdot{\bf x}\pm \omega t)}.
\end{eqnarray}
\end{mathletters}
where ${\bf k}$ and $\omega$ are constants, and $a_{\nu}$, $b_{\nu}$ and $c_{\nu}$ are polarization vectors.
For a linearly polarized wave, described by Eqs.\ (\ref{eq-photonresult}),
the interaction terms ${\cal L}_{int}$ and ${\cal L}_{int+}$
cancel, simplifying the equations of motion. Substituting Eqs.\ (\ref{eq-photonresult})
into the equations of motion (\ref{eq-vectoreom}) and
solving for $k$ results in the following polynomial in $\omega$:
\begin{eqnarray}
\left\{\left[\omega^{2}-k^2-\rho^{2}\sin^{2}\theta_{W}(1-\delta_{\nu z})\right]\right. \nonumber \\
\times \left.\left[\omega^{2}-k^{2}-M_{W}^{2}-\rho^{2}(1-\delta_{\nu z})\right]+4 \rho^{2} k_{z}^{2}(1-\delta_{\nu z})\right\} \nonumber \\
\times \left\{\left[\omega^{2}-k^{2}-M_{Z}^{2}-\rho^{2}\cos^{2}\theta_{W}(1-\delta_{\nu z})\right]\right. \nonumber \\
\times \left.\left[\omega^{2}-k^{2}-M_{W}^{2}-\rho^{2}(1-\delta_{\nu z})\right]+4\rho^{2} k_{z}^{2}(1-\delta_{\nu z})\right\} \nonumber \\
= \left\{\rho^{2}\sin^{2}\theta_{W} \left[\omega^{2}-k^{2}-M_{W}^{2}-\rho^{2}(1-\delta_{\nu z})\right] + 4\rho^{2}k_{z}^{2}(1-\delta_{\nu z})\right\} \nonumber \\
\label{eq-krestrict}
\times \left\{\rho^{2}\cos^{2}\theta_{W} \left[\omega^{2}-k^{2}-M_{W}^{2}-\rho^{2}(1-\delta_{\nu z})\right] + 4\rho^{2}k_{z}^{2}(1-\delta_{\nu z})\right\} .
\end{eqnarray}
Here $k=|{\bf k}|$ and $k_{z}$ denotes the component of ${\bf k}$ in the $z$-direction.
Equation (\ref{eq-krestrict}) governs the relationship between $k$ and
$\omega$ and hence determines the propagation speed of the vector bosons.
Note that $\delta_{\nu z}\neq0$ when the polarization direction of the
gauge bosons has a component in the direction of isospin rotation; hence different polarization
states propagate at different velocities.

The equations governing the field intensities can also be written in
terms of the photon field intensity $a_{\nu}$, i.e.,
\begin{mathletters}
\label{eq-abcsoln}
\begin{eqnarray}
b_{\nu}\{\omega^{2}-k^{2}-\rho^{2}[(\cos\theta_{W})^{2}-\sin^{2}(\theta_{W})^{2}]-M_{Z}^{2}\} \nonumber \\
= a_{\nu} \{\omega^{2}-k^{2}-\rho^{2}[(\sin\theta_{W})^{2}-(\cos\theta_{W})^{2}]\} \\
\label{eq-wgoven}
c_{\nu}\left(\omega^{2}-k^{2}-\rho^{2}-M_{W}^{2}\right) = (a_{\nu}+b_{\nu})(2ik_{z}\rho).
\end{eqnarray}
\end{mathletters}
The $\alpha^{W}_{\nu}$-field decouples from the photon and $Z^{0}$
fields when there is no component of $k$ in the $z$-direction (i.e., $k_{z}=0$).
However, for $k_{z}\neq 0$ the
$\alpha^{W}_{\nu}$-field is coupled to, and $90^{o}$ out of phase with the
photon and $Z^{0}$ fields. This arises because of the imaginary term in Eq.\ (\ref{eq-wgoven}).

When the isospin rotation strength is zero ($\rho=0$), Eq.\ (\ref{eq-krestrict})
simplifies to
\begin{equation}
\label{eq-original}
(\omega^{2}-k^{2})(\omega^{2}-k^{2}-M_{W}^{2})(\omega^{2}-k^{2}-M_{Z}^{2})=0.
\end{equation}
The solutions to Eq.\ (\ref{eq-original}) correspond to the massless photon ($\omega=\pm k$)
and massive $Z^{0}$ and $W^{\pm}$ intermediate vector bosons
($E^{2}-p^2=m^{2}$). For $\rho=0$,
the solutions to Eq.\ (\ref{eq-original}) separate into the distinct fields,
$\alpha^{\gamma}_{\nu}$, $\alpha^{Z}_{\nu}$ and $\alpha^{W}_{\nu}$.
This is because in this case the solutions for $a_{\nu}$, $b_{\nu}$ and $c_{\nu}$
decouple in Eq.\ (\ref{eq-abcsoln}).
Hence each vector boson field can be determined independently of the
others. However, when $\rho\neq0$, the vector bosons are coupled and it is
no longer possible to find independent solutions for the individual fields
(i.e., $a_{\nu}$, $b_{\nu}$ and $c_{\nu}$ are all non-zero for a given
solution). There
are still three solutions corresponding to three propagating states, however,
each of these three solutions consist of a mixture of intermediate
vector bosons. As $\rho$ is reduced to zero, these mixed solutions
deform continuously into independent solutions for the spin-1 fields. Consequently,
for small isospin rotation, we can interpret the mixed
particle solutions as describing predominantly photons, $Z^{0}$ and $W^{\pm}$ particles.

For the case $\rho \ll k \ll |\Phi|$, the photon
solution to Eq.\ (\ref{eq-krestrict}) may be written as
\begin{equation}
\label{eq-ppmass}
\omega^2=k^2+\rho^{2}\sin^{2}\theta_{W}(1-\delta_{\nu z})+{\cal O}\left(\frac{\rho^{2} k_{z}^2}{M_{W}^{2}}\right)+{\cal O}\left(\frac{\rho^{4}}{|\Phi|^{2}}\right).
\end{equation}
Equation (\ref{eq-ppmass}) implies that the photon propagates as if it had a
pseudo-mass $P_{\gamma}$, whence
Eq.\ (\ref{eq-pseudom}) becomes
\begin{equation}
P_{\gamma}\approx \rho\sin\theta_{W}.
\end{equation}
However, the photon has zero rest mass since its mass matrix eigenvalue is
always zero; nevertheless, it propagates with a phase velocity
less than the speed of light for a constant Higgs field.
When the isospin rotation strength, $\rho$, is small we can model the
behaviour of the photon using an effective refractive index $n_{H}$.
For $\rho=const$ we find (for $\omega \gg P_{\gamma}$)
\begin{equation}
\label{eq-refractive}
n_{H} \equiv \frac{k}{\omega} \approx \sqrt{1-\frac{P_{\gamma}^{2}}{\omega^{2}}} \approx 1 -\frac{1}{2}\frac{P_{\gamma}^{2}}{\omega^{2}}+{\cal O}(P^{4}_{\gamma}/\omega^{4}).
\end{equation}
The mixed particle solution, arising from the isospin coupling,
only propagates with $k < \omega$; this is due to the admixture of massive and
massless spin-1 bosons. When $\omega=0$, the equations describe a finite
range electromagnetic field, where $k=\pm iP_{\gamma}$.

The refractive index (\ref{eq-refractive}) depends
on whether the polarization direction is parallel, or perpendicular to
the rotation of the Higgs field in isospin space. Hence we can write
\begin{mathletters}
\begin{eqnarray}
n_{\bot} & \approx & 1-\frac{1}{2}\frac{P^{2}_{\gamma}}{\omega^{2}}+{\cal O}(P_{\gamma}^{4}/\omega^{4}) \\
n_{\|} & = & 1.
\end{eqnarray}
\end{mathletters}
Therefore a region where the isospin orientation of the
Higgs field varies spatially acts as a birefringent medium \cite{born52}.
The relative phase shift, $\Delta \varphi$, between orthogonal polarization states
propagating through the birefringent medium is given by
\begin{equation}
\label{eq-varphi}
\Delta\varphi = k d |n_{\bot}-n_{\|}|,
\end{equation}
where to first order the birefringence length scale is ($k \gg P_{\gamma}$)
\begin{equation}
\label{eq-bilen}
d \approx \frac{4\pi k}{P_{\gamma}^{2}}.
\end{equation}

Equation (\ref{eq-varphi}) can be generalised to
\begin{equation}
\Delta\varphi = d P_{\gamma}^{2} \sin\theta/ 2 k,
\end{equation}
where $\theta$ is the angle between the direction of propagation and
the vector $\rho_{\mu}$. The angular dependence of the phase shift arises because
an electromagnetic wave propagating parallel to $\rho_{\mu}$ has both polarization
vectors perpendicular to $\rho_{\mu}$ (i.e., no relative phase shift); compared
to a wave propagating perpendicular to $\rho_{\mu}$, in which one polarization
state is parallel, and the other perpendicular to $\rho_{\mu}$ (i.e., a phase
shift of $\Delta \varphi = d P^{2}_{\gamma}/ 2 k$).

\section{Coupling Strength For Isospin Rotation}
\label{sec-measure}
To calculate the size of $P_{\gamma}$ and $d$ we require an estimate of the isospin
rotation of the Higgs field. We can use a homilia string network
\cite{thatcher99} to estimate $\rho_{\mu}$, since this defect is stable in the
electroweak model. However, birefringent properties arise for any defect which predicts
an isospin rotation of the Higgs field.

Homilia strings induce an isospin rotation at all points in the Universe, since the functions,
$|\phi^{\alpha}(r)|$ and $|\phi^{\beta}(r)|$, never
limit to a constant value (see e.g., the vortex solution in reference \cite{thatcher99}).
This behaviour reflects the texture nature of the
homilia string network.
In \cite{thatcher99} we obtained an approximate vortex
solution for a homilia $\alpha$-string based on cylindrical symmetry.
We describe the homilia string network by a single length
scale, $L(t)$, in the same manner as cosmic string networks are characterized
\cite{martin96}. Here $L(t)$ is the average distance between
a segment of homilia string and its nearest neighbours. The vortex solution
is invariant under rescaling,
${\bf x}\rightarrow {\bf x}'= {\bf x}/L(t)$.
Hence we write the homilia $\alpha$-string for an arbitrary length scale $L(t)$ as
\begin{equation}
\label{eq-scaledvortex}
\Phi(r')=\left(\begin{array}{c} |\phi^{\alpha}(r')| e^{i\theta} \\ |\phi^{\beta}(r')| \end{array}\right).
\end{equation}
This vortex solution is scale invariant, since
for $L \gg (\sqrt{\lambda}\eta)^{-1}$ we find
$|\Phi|\approx const$. Note that in general
$|\Phi|\neq const$ for the homilia string vortex solution since there
must always be a deviation from $\eta$ at $r=0$ \cite{thatcher99}. However,
the size of this deviation depends on the separation distance and
for large separations it is a good approximation to write $|\Phi|\approx const$.
The energy density of the homilia string network increases as we reduce
$L$ and hence the invariance of the vortex solution under rescaling by $L$
does not result in the collapse of the network.

Using Eq.\ (\ref{eq-scaledvortex}) we find
\begin{equation}
\label{eq-pmass}
P_{\gamma}(t)=\sin\theta_{W}(\rho_{\mu}(t)\rho^{\mu}(t))^{1/2} \approx\sin\theta_{W} \frac{(\hat{\rho}_{\mu}\hat{\rho}^{\mu})^{1/2}}{ L(t)},
\end{equation}
where $\hat{\rho}_{\mu}\hat{\rho}^{\mu}$ is a numerically determined quantity
that describes the strength of the isospin rotation of the homilia string. 
From a cylindrically symmetric $SU(2)\times U(1)$
homilia string \cite{thatcher99}, we obtain an average value of
$<\hat{\rho}_{\mu}\hat{\rho}^{\mu}>^{1/2}\approx 1.3$,
with $L(t)=1\mbox{GeV}^{-1}$. Equation (\ref{eq-pmass})
describes the relationship between the pseudo-mass of the photon
and the length scale of the homilia string network.

The quantity $L(t)$ is a measure of the
length scale of the network and can be estimated from
analytical models of string network evolution \cite{martin96}. To take into
account the two types of electroweak homilia strings (HS) we write
\begin{equation}
L(t) = L_{HS}(t) \approx \frac{1}{\sqrt{N}} L_{CS}(t),
\end{equation}
where $N$ is the order of the symmetry breaking group, $U(N)$,
and $L_{CS}$ denotes the length scale for a cosmic string network. For the
electroweak model $N=2$. Consider a flat Universe with a zero
cosmological constant. $L(t)$ can be approximated in the
matter dominated era by \cite{martin96}
\begin{equation}
L(t) = \left(\frac{9 k_{m}(k_{m} +c_{m})}{8 N}\right)^{1/2}t,
\end{equation}
where $k_{m} = 0.49$ describes the small scale structure and $c_{m}=0.17$
is the loop chopping efficiency \cite{martin96}. For electroweak homilia strings we
find $L(t) \approx 0.43 t$, hence the length scale of the photon pseudo-mass
is of the order of the size of the observable Universe.

Although the photon pseudo-mass is very small, it can be differentiated
from quantum fluctuations since the former is coherent over the length scale $L(t)$.
Because the length scale
is $L(t) \approx 0.43 t$, the direction of $\rho_{\mu}$ is essentially
constant across the observable Universe.
As a consequence the birefringence scale
exhibits an angular dependence described by $d \propto k / (\sin\theta P_{\gamma}^{2})$,
where $\theta$ is the angle between the direction of propagation and
the vector $\rho_{\mu}$.

Isospin rotation can in principle be measured by examining the orientation of
polarized synchrotron radiation emitted from distant active galactic nuclei.
Using the estimate from Eq.\ (\ref{eq-bilen}) we find the birefringence
length scale of the homilia network to be
\begin{equation}
d \approx \frac{\omega (0.43 t)^{2}}{(\hat{\rho}\sin\theta_{W})^{2}}.
\end{equation}
The typical frequency of synchrotron radiation from distant galaxies is
$\omega \sim 10^{-16}$ GeV and hence in the present epoch we obtain an
order of magnitude estimate of $d \sim 10^{65}$ $\mbox{GeV}^{-1}$, 
which is $10^{24}$ times the size of the
observable Universe. Nodland and Ralston \cite{nodland97} claim
to have detected a birefringence length
scale of the order of the size of the observable Universe.
Therefore, the isospin rotation of the Higgs field due to the presence
of a homilia string network does not explain their observation \cite{nodland97}.
Although we could conjecture a smaller network length scale, this would not
explain why the orientation of the isospin rotation of the Higgs field
is coherent over ten billion lightyears.

For the model to be in agreement with
the results of Nodland and Ralston \cite{nodland97}, we require
the existence of an electroweak texture which possesses a
$\rho_{\mu}$-direction which is coherent across the observable Universe and
induces an average pseudo-mass of
$<P_{\gamma}>\approx 10^{-28} \mbox{GeV}^{-1}=10^{-52}$ g.
The average pseudo-mass of $10^{-52}$ g is two orders of magnitude
smaller than the upper limit of the `photon mass' as measured by Lakes
\cite{lakes98}. It is conceivable that the existence of such a
texture could be verified in the near future.

\section{Conclusions}
In this paper we have examined the behaviour of the photon field in
an electroweak homilia string network. It is found that the spatial dependence of
the orientation of the Higgs field
in isospin space results in additional couplings between
the photon and intermediate vector bosons. This gives rise to a
photon pseudo-mass which depends on the polarization state of the photon; 
hence an homilia string network acts like a birefringent medium.
Isospin rotation of the Higgs field can be differentiated from quantum
effects, since the former is coherent over a
length scale of the order of the observable Universe. However, there is
a large discrepancy between the predicted birefringent length scale and the
observed length scale reported in reference \cite{nodland97}.
It will require further investigation in order to determine whether the
observed birefringence scale can be reconciled within the standard electroweak
model.

\acknowledgments
One of the authors (M.J.T.) acknowledges the support of the APA.

\appendix
\section{Gauge Invariant Eigenvectors}
\label{sec-gauge}
The vector boson eigenvectors (\ref{eq-gaugeeigenvectors}) are valid for
a particular choice of local gauge in which the Higgs field is real.
If the local gauge is not fixed the particle eigenvectors become
\begin{mathletters}
\label{eq-pgidefn}
\begin{eqnarray}
{\bf A}^{\gamma}(x) & = &\frac{\sin\theta_{W}}{\tan\theta_{W}|\Phi|^{2}} \left(\begin{array}{c} -|\Phi|^{2} \\ \chi^{1}(x) \\ \chi^{2}(x) \\ \chi^{3}(x) \end{array}\right) \\
{\bf A}^{Z}(x) & = &\frac{\cos\theta_{W}}{\tan\theta_{W}|\Phi|^{2}} \left(\begin{array}{c} |\Phi|^{2}\tan^{2}\theta_{W} \\ \chi^{1}(x) \\ \chi^{2}(x) \\ \chi^{3}(x) \end{array}\right) \\
\label{eq-wgidefn}
{\cal N}_{\mu}(x){\bf A}^{W}(x) & = &\left(\begin{array}{c} 0\\ {\cal A}_{\mu}(x)\chi^{2}(x)+{\cal B}_{\mu}(x)\chi^{3}(x) \\ -{\cal A}_{\mu}(x)\chi^{1}(x)+{\cal C}_{\mu}(x)\chi^{3}(x) \\
-{\cal B}_{\mu}(x)\chi^{1}(x)-{\cal C}_{\mu}(x)\chi^{2}(x) \end{array}\right) \\
\label{eq-owgidefn}
{\cal N}_{\mu}(x){\bf A}^{\overline{W}} & = & i|\Phi|^{2}\tan\theta_{W}\left(\begin{array}{c} 0 \\ -{\cal C}_{\mu}(x) \\ {\cal B}_{\mu}(x) \\ -{\cal A}_{\mu}(x) \end{array}\right),
\end{eqnarray}
\end{mathletters}
where
\begin{mathletters}
\begin{eqnarray}
{\cal A}_{\mu} & = & \frac{\chi^{2}\partial_{\mu}\chi^{1}-\chi^{1}\partial_{\mu}\chi^{2}}{|\Phi|^{4}\tan^{2}\theta_{W}} \\
{\cal B}_{\mu} & = & \frac{\chi^{3}\partial_{\mu}\chi^{1}-\chi^{1}\partial_{\mu}\chi^{3}}{|\Phi|^{4}\tan^{2}\theta_{W}} \\
{\cal C}_{\mu} & = & \frac{\chi^{3}\partial_{\mu}\chi^{2}-\chi^{2}\partial_{\mu}\chi^{3}}{|\Phi|^{4}\tan^{2}\theta_{W}}.
\end{eqnarray}
\end{mathletters}
The normalisation constant ${\cal N}_{\mu}$ is
\begin{equation}
\label{eq-norm}
{\cal N}_{\mu} = \frac{\sqrt{(\partial^{\mu}\chi^{a})(\partial_{\mu}\chi_{a})}}{|\Phi|^{2}\tan\theta_{W}},
\end{equation}
where the summation convention does not apply to $\mu$ in Eq.\ (\ref{eq-norm}).
Note the eigenvectors for $W$ and $\overline{W}$ (\ref{eq-wgidefn} and \ref{eq-owgidefn})
cannot be expressed independently of the vector coupling fields
${\cal A}_{\mu}$, ${\cal B}_{\mu}$ and ${\cal C}_{\mu}$.

From the definitions of the vector boson eigenvectors in Eqs.\ (\ref{eq-pgidefn})
we derive the relations
\begin{mathletters}
\label{eq-gencoup}
\begin{eqnarray}
(\sin\theta_{W})^{-1}\partial_{\mu} {\bf A}^{\gamma} & = & (\cos\theta_{W})^{-1}\partial_{\mu}{\bf A}^{Z} = {\cal N}_{\nu} {\bf A}^{W} \\
\partial_{\mu}{\bf A}^{W} & = & {\cal N}_{\nu} ({\bf A}^{\gamma}\sin\theta_{W}+{\bf A}^{Z}\cos\theta_{W})+\tau_{\mu}{\bf A}^{\overline{W}} \\
\partial_{\mu}{\bf A}^{\overline{W}} & = & -\tau_{\mu}{\bf A}^{W}.
\end{eqnarray}
\end{mathletters}
Comparing Eqs.\ (\ref{eq-gencoup}) with Eqs.\ (\ref{eq-photonresult}) shows that
the coupling strength $\rho_{\mu}$ is given by
\begin{equation}
\label{eq-newrhob}
\rho_{\mu} \equiv {\cal N}_{\mu} =\frac{\sqrt{(\partial_{\mu}\chi^{a})(\partial^{\mu}\chi_{a})}}{|\Phi|^{2}\tan\theta_{W}},
\end{equation}
where $a=1$, 2 and 3, and
the summation convention does not apply to $\mu$. Using Eq.\ (\ref{eq-newrhob}),
$\rho_{\mu}\rho^{\mu}$ is of the form
\begin{equation}
\label{eq-newrho}
\rho_{\mu}\rho^{\mu} = \frac{(\partial_{\mu}\chi^{a})(\partial^{\mu}\chi_{a})}{|\Phi|^{4}\tan^{2}\theta_{W}}.
\end{equation}
$\tau_{\mu}$ is defined as
\begin{equation}
N_{\alpha}N_{\beta}\tau_{\delta} = \frac{\epsilon_{abc}\chi^{a}(\partial_{\alpha}\chi^{b})\partial_{\beta}\partial_{\delta}\chi^{c}}{|\Phi|^{2}\tan\theta_{W}},
\end{equation}
where $\epsilon_{abc}$ is the fully asymmetric tensor and we sum over $\alpha$ and $\beta$.
In Eqs.\ (\ref{eq-fcoup}) $\tau_{\mu}=0$ due to the choice of gauge which
resulted in $\chi^{2}=0$. We can always perform a gauge transformation
such that $\tau_{\mu}=0$ and hence $\tau_{\mu}$ does not described a physical
coupling.
Employing the relations in Eqs.\ (\ref{eq-gencoup}) leads to the field strength $Tr(F_{\mu\nu}F^{\mu\nu})$
being described by Eq.\ (\ref{eq-ewlagfullterms}), with $\rho_{\mu}$ defined
by Eq.\ (\ref{eq-newrhob}).

\end{document}